\documentclass[amssymb,amsmath,aps,groupedaddress,reprint,superscriptaddress,twocolumn,showpacs]{revtex4-1}

\usepackage{hyperref}
\usepackage{graphicx}
\usepackage{dcolumn}
\usepackage{bm}
\usepackage{amsmath}

\begin{document}
\title{Laser-driven parametric instability and generation of entangled photon-plasmon states in graphene and topological insulators}
\author{Mikhail Tokman}
\affiliation{Institute of Applied Physics, Russian Academy of Sciences}
\author{Yongrui Wang}
\affiliation{Department of Physics and Astronomy, Texas A\&M
University, College Station, TX, 77843 USA}
\author{Ivan Oladyshkin}
\affiliation{Institute of Applied Physics, Russian Academy of Sciences}
\author{A. Ryan Kutayiah}
\affiliation{Department of Physics and Astronomy, Texas A\&M
University, College Station, TX, 77843 USA}
\author{Alexey Belyanin}
\affiliation{Department of Physics and Astronomy, Texas A\&M
University, College Station, TX, 77843 USA}

\date{\today}

\begin{abstract}

We show that a strong infrared laser beam obliquely incident on graphene can experience a parametric instability with respect to decay into lower-frequency (idler) photons and THz surface plasmons. The instability is due to a strong in-plane second-order nonlinear response of graphene which originates from its spatial dispersion. The parametric decay leads to efficient generation of THz plasmons and gives rise to quantum entanglement of idler photons and surface plasmon states. A similar process can be supported by surface states of topological insulators such as Bi$_2$Se$_3$.
\end{abstract}


\maketitle

\section{Introduction}

Nonlinear parametric decay of a pump laser photon into two lower-frequency photons (usually called "signal" and "idler") in a nonlinear crystal possessing a second-order nonlinearity is the most popular method of generating entangled photon states  \cite{kwiat}. At higher pump intensities the parametric process can experience gain which leads to the instability and exponential amplification of coupled signal and idler fields. Stimulated parametric decay enables optical parametric amplifiers and oscillators as popular tunable sources of long-wavelength radiation from near- to far-infrared \cite{shen}. They typically employ bulk transparent crystals under phase-matching conditions for frequencies and wave vectors of the fields participating in a three-wave mixing interaction: 
\begin{align}
&\omega_{s}=\omega_{p}-\omega_{i} ; \; 
\bm{k}_{s}=\bm{k}_{p} - \bm{k}_{i}, \label{1}
\end{align}
where the subscripts $s$, $p$, and $i$ represent signal, pump and idler, respectively. 
In view of these requirements, the very idea of parametric amplification supported by just a monolayer of material seems unrealistic. Surprisingly,  we find that stimulated parametric decay of laser photons is feasible in 2D systems of massless Dirac electrons. 

Any surface has anisotropy between in-plane and out-of-plane excitations, and graphene is no exception. However, the  second-order susceptibility $\chi^{(2)}_{ijk}$ associated with this surface anisotropy is very small in graphene \cite{vandriel} and we don't even consider it below. A much stronger nonlinear response is expected when all fields and electron excitations lie in-plane. This is obvious already in the classical free-carrier limit because of an extreme band nonparabolicity \cite{mikhailov2008}. However, graphene is a centrosymmetric system for low-energy in-plane excitations, which should prohibit any second-order response. Nevertheless, a non-zero $\chi^{(2)}$ appears beyond the electric dipole approximation when one includes the  dependence of $\chi^{(2)}$ on the in-plane photon wave vectors, i.e. the spatial dispersion. In this case the isotropy of graphene is effectively broken by the wave vector direction. The spatial dispersion effects turn out to be quite large because of a large magnitude of the electron velocity $v_F$, similarly to spatial dispersion in a hot plasma. Further enhancement of $\chi^{(2)}$ occurs at resonance between the pump frequency and twice the Fermi energy: $\omega_p = 2 \epsilon_F/\hbar$. Finally, the efficiency of parametric down-conversion is enhanced when one of the generated fields is not a photon but a surface plasmon mode supported by a massless 2D electron layer. A non-zero value of the nonlocal in-plane $\chi^{(2)}$ and plasmon enhancement of the nonlinear signal were pointed out before for second-harmonic generation \cite{mikhailovSHG, smirnova2014} (which only included intraband transitions in a free-carrier model) and for difference-frequency generation \cite{5}. Here we develop the first theory of the parametric decay in graphene, which includes fully quantum description of the nonlinear response and quantization of all fields.  

\begin{figure}[ht]
  \centering
  \includegraphics[height=5cm,width=8.5cm]{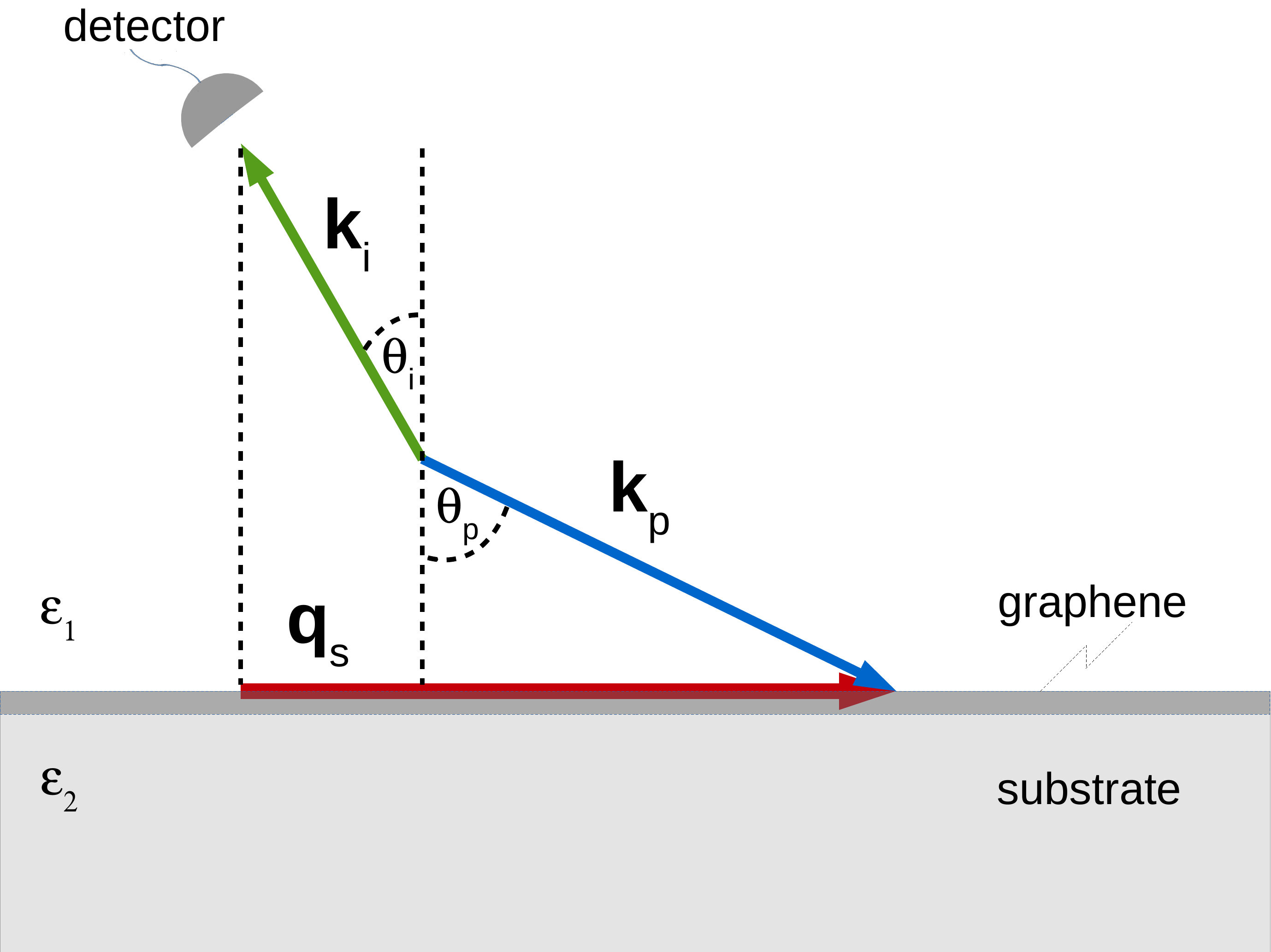}\llap{\raisebox{2cm}{\includegraphics[height=2.7cm]{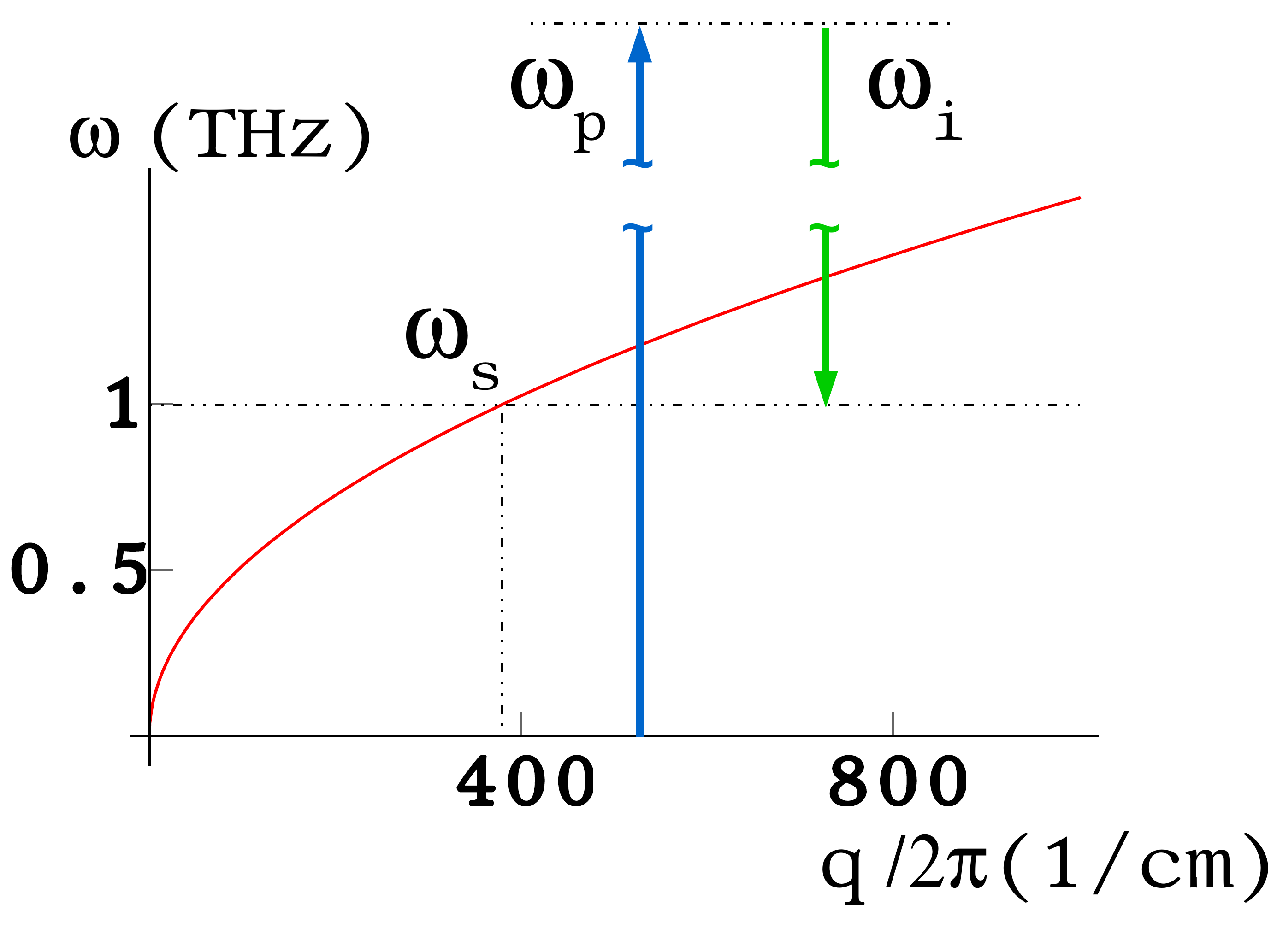}}} \\
  \caption{Schematic of the parametric decay of the pump photon into an idler photon and a surface plasmon, which satisfies conservation of energy and in-plane component of momentum. Inset shows matching of the signal frequency to the real part of surface plasmon frequency obtained by solving Eq.~(\ref{5}).}
\end{figure}

The schematic of the nonlinear process is shown in Fig.~1. An obliquely incident pump photon decays into an idler photon and a "signal"  plasmon of a much lower frequency $\omega_{s}=\omega_{p}-\omega_{i} \ll \omega_{p,i}$ but a comparable wave vector $q_s \sim  q_p$. The second of phase matching conditions in Eqs.~(\ref{1}) is replaced by its in-plane projection $\bm{q}_{s}=\bm{q}_{p} - \bm{q}_{i}$. In addition, the signal frequency should match the real part of surface plasmon dispersion $\omega(q)$ shown in the inset to Fig.~1: $\omega_s = \omega(q_s)$. Note that both positive and negative projections of the idler wave vector $q_i$ are possible, where the positive direction is assumed to the right. In particular, negative values of $q_i$ give access to larger plasmon wave vectors $q_s = q_p - q_i = |q_p| + |q_i|$ and frequencies.   

\section{Quantized surface plasmon field in graphene}
 
Consider a geometry of Fig.~1, i.e. a 2D layer of massless Dirac electrons in $z = 0$ plane between two  media with dielectric constants $\epsilon_1$ and $\epsilon_2$.   The plasmon frequency $\omega_{s}$ and in-plane wave vector $\bm{q}_{s}$ are related through the dispersion relation for a TM-polarized surface mode \cite{5}:
\begin{equation}
\label{5}
4\pi\chi_{s}+\frac{\epsilon_{1}}{p_1}+\frac{\epsilon_{2}}{p_2}=0,
\end{equation}
where $p_{1,2}=\sqrt{q_{s}^{2}-\epsilon_{1,2}\frac{\omega_{s}^{2}}{c^{2}}}$. At the THz frequencies smaller than twice the Fermi energy $2\epsilon_{F}/\hbar$ one needs only to take into account the intraband contributions to the linear 2D susceptibility $\chi_{s}(\omega_{s},q_{s})$, which in the limit of strong degeneracy is given by \cite{5} 
\begin{align}
& \chi_{s}(\omega_{s},q_{s})=\frac{2e^{2}E_{F}}{\pi \hbar^{2}\omega_{s}}\frac{(\omega_{s}+i\gamma_s)}{(v_{F}q_{s})^{2}}  \times 
\left[1-  \right. \nonumber \\  
& \left. \frac{\omega_{s}+i\gamma_s}{\omega_{s}+i\gamma+v_{F}q_{s}}\sqrt{1+\frac{2v_{F}q_{s}}{\omega_{s}+i\gamma_s-v_{F}q_{s}}}\right] \label{chilinear}
\end{align}
where $E_{F}$ is the Fermi energy and $\gamma_s$ is the decay rate of surface plasmons.  

The standard quantization procedure in the limit of $\omega_s \gg \gamma$ leads to 
\begin{equation}
\label{4}
\bm{\hat{E}}=\sum_{\bm{q}_{s}} \bm{E}_{s}(z)\hat{a}_s e^{i\bm{q}_{s}\bm{r}_{\|}-i\omega_{s}t}+{\rm H.c.}
\end{equation}
where  $\bm{r}_{\|}=(x,y)$ and $\hat{a}_s, \hat{a}^{\dagger}_s$ are annihilation and creation operators of surface plasmon modes. The $z$-distribution of the field $\bm{E}_{s}(z)$ is  \cite{5}:
\begin{align}
&\bm{E}_{s}(z)=\left( \bm{x}_{0}\pm\bm{z}_{0}\frac{iq_{s}}{p_{1,2}} \right) E_{s0}e^{\mp p_{1,2}z}, \label{7} 
\end{align}
where the upper and lower signs correspond to $z>0$, $z<0$, respectively. 
Similarly to the case of propagating fields  \cite{8,19}, the energy of the plasmon field inside a volume $V$ can be written as
\begin{equation}
\nonumber
\hat{H}=\frac{1}{8\pi}\sum_{s}(\hat{a}_{s}^{\dagger}\hat{a}_{s}+\hat{a}_{s}\hat{a}_{s}^{\dagger}) \int_{V}\left(\bm{E}_{s}^{*}\frac{\partial (\omega\tilde{\epsilon})}{\partial \omega}\bm{E}_{s}+\bm{B}_{s}\bm{B}_{s}^{*}\right)d^{3}r,
\end{equation}
where $\tilde{\epsilon}$ is the dielectric permittivity tensor and the projection of the quantization volume onto the 2D layer is equal to 1. 
After substituting Eqs.~(\ref{5}) and (\ref{7}), the last equation is reduced to a standard Hamiltonian for the plasmon field, $\hat{H}_s = \sum_{\bm{q}_{s}} \hbar \omega_s(\bm{q}_{s}) \left( \hat{a}^{\dagger}_s  \hat{a}_s + 1/2 \right)$, if we choose the normalization  constant $E_{s0}$ (in quasi-electrostatic approximation) as:  
\begin{equation}
\label{9}
\left|E_{s0}\right|^{2}=\hbar\left({\rm Re}(\partial \chi_{s}/\partial \omega)\right)^{-1}.
\end{equation}

The effect of dissipation of a plasmon field (within $\omega_s \gg \gamma_s$) and its nonlinear interaction with other fields can be taken into account within the Heisenberg-Langevin approach \cite{19}. For quasi-monochromatic wave fields, it is convenient to consider a wave packet of surface plasmon modes with frequencies and wave vectors concentrated in a narrow spectral range $\Delta \omega \ll \omega_s$, $\Delta q \ll q_s$ near a central component $\propto e^{i\bm{q}_{s}\bm{r}_{\|}-i\omega_{s}t}$ \cite{10,11,13,14}.  Within this approach we introduce the annihilation and creation operators $\hat{a}_{s}(\bm{r}_{\|},t)$ and $\hat{a}_{s}^{\dagger}(\bm{r}_{\|},t)$ that are slowly varying in time and space relative to $\omega_s$ and $q_{s}$. Their commutator is equal to the number of quantized modes per unit area $L_x \times L_y = 1$ within the spectral interval $\Delta \omega$: $[\hat{a}_{s},\hat{a}_{s}^{\dagger}]= \displaystyle \frac{\Delta \omega}{2\pi L_{y}v_{s}}$,
where $v_{s} $ is the group velocity of a surface plasmon which determines its spectral density of states  and $L_{y}$ is the aperture size of the beam.

Equations for a slowly varying field amplitude of a surface plasmon wave packet can be obtained in the same way as for the propagating optical fields; see e.g. \cite{11,13,14}:
\begin{equation}
\label{15}
\frac{\partial \hat{a}_{s}}{\partial t}+v_{s}\frac{\partial \hat{a}_{s}}{\partial x}+\gamma_{s}\hat{a}_{s}=\frac{i}{\hbar}\hat{P}_{s}^{(2)}E_{s0}^{*} + \hat{F}_{s},
\end{equation}
where
$\gamma_{s} = \hbar^{-1} ({\rm Im}[\chi_s])|E_{s0}^2|$, $\hat{F}_{s}(t)$ is the operator of the Langevin noise, and $\hat{\bm{P}}_s^{(2)}=\bm{x}_{0}\hat{P}_{s}^{(2)}e^{iq_{s}x-i\omega_{s}t}+$ H.c. is the second-order nonlinear component of the operator of the polarization.  

The Langevin noise source ensures a correct expression for the commutator of the plasmon field in the presence of its  interaction  with a dissipative reservoir. It is convenient to define the properties of the noise source in terms of its spectral components  $\hat{F}_{s}=\int \hat{F}_{s\omega} e^{-i\omega t} \, d\omega$ and $\hat{F}_{s\omega}^{\dagger}=\hat{F}_{s;-\omega} $. Assuming a dissipative reservoir in thermal equilibrium and adjusting for the 2D geometry, we can write \cite{19}
\begin{align}
\label{46}
\left<\hat{F}^{\dagger}_{\omega^{'}}(x')\hat{F}_{\omega}(x)\right> & = \frac{\gamma_{s}N_{T}(\omega_{s})}{\pi L_{y}}\delta(\omega-\omega^{'})\delta(x-x^{'});  \\
\left<\hat{F}_{\omega}(x) \hat{F}^{\dagger}_{\omega^{'}}(x') \right> & = \frac{\gamma_{s}(N_{T}\left(\omega_{s})+1\right)}{\pi L_{y}}\delta(\omega-\omega^{'})\delta(x-x^{'}), \nonumber
\end{align}
where the $\left< ... \right>$ means averaging over both an initial quantum state in the Heisenberg picture and the statistics of the dissipative reservoir, 
$N_{T}(\omega_{s})=\left({\rm e}^{\hbar\omega_{s}/(k_B T)}-1\right)^{-1}$.
In the absence of the nonlinear polarization, the solution of Eqs.~(\ref{15}),(\ref{46}) in the limit $\gamma_s x/v_s \rightarrow \infty$ corresponds to thermal equilibrium: 
\begin{equation}
\label{22}
\left<\hat{a}_{s}^{\dagger}\hat{a}_{s}\right> \rightarrow \left<\hat{a}_{s}^{\dagger}\hat{a}_{s}\right>_{T}=\frac{N_{T}(\omega_{s}) \Delta\omega}{2\pi L_y v_s}.
\end{equation}
Eq.~(\ref{22}) corresponds to a general property of thermal emission: its power received by a matched antenna $ \approx L_y v_s \hbar \omega_s \left<\hat{a}_{s}^{\dagger}\hat{a}_{s}\right>_{T}$ does not depend on the size and shape of the aperture \cite{Bekefi}. \\

\section{Parametric instability in graphene} 

Consider parametric second-order interaction of surface plasmons with an s-polarized bichromatic pump+idler field incident from the $z > 0$ half-space:  
$$ \hat{\bm{E}} = \sum_{j = p,i}  \hat{\bm{E}}_j; \, \hat{\bm{E}}_j = \bm{y}_0 E_{j0} \hat{c}_j {\rm e}^{-ik_j z + i q_j x - i \omega_j t} + {\rm H.c.}, 
$$
where the normalization fields $|E_{j0}|^2 = 2 \pi \hbar \omega_j/n_1^2$ are defined for a unit quantization volume, $\hat{c}_j$ are Heisenberg operators of slowly varying amplitudes corresponding to a finite spectral width $\Delta \omega$ \cite{11,13,14,17}. The nonlinear 2D polarization at frequencies $\omega_{p,i,s}$ generated in the graphene plane $z=0$ is given by 
\begin{align}
& \hat{\bm{P}}_s^{(2)} = \bm{x}_0 \chi_{xyy}^{(s,2)} \hat{\cal{E}}_i^{\dagger} \hat{\cal{E}}_p + {\rm H.c.}, \nonumber \\
& \hat{\bm{P}}_i^{(2)} = \bm{y}_0 \chi_{yyx}^{(i,2)} E_{s0}^* \hat{a}_s^{\dagger} \hat{\cal{E}}_p {\rm e}^{-i q_s x + i \omega_s t} + {\rm H.c.}, 
\nonumber \\
&  \hat{\bm{P}}_p^{(2)} = \bm{y}_0 \chi_{yyx}^{(p,2)} E_{s0} \hat{a}_s \hat{\cal{E}}_i {\rm e}^{i q_s x - i \omega_s t} +  {\rm H.c.}, \label{polar}
\end{align}
where $\hat{\cal{E}}_{p,i}$ are the $\propto \exp(-i\omega t)$ parts of the electric field operators at the pump and idler frequencies $\omega_{p,i}$ in the graphene plane. They are related to the field operators in the upper and lower half-spaces through standard boundary conditions.  The second-order susceptibilities at corresponding frequencies are $\chi_{xyy}^{(s,2)} = \chi_{xyy}^{(2)}(\omega_s = \omega_p - \omega_i)$,  $\chi_{yyx}^{(i,2)} = \chi_{yyx}^{(2)}(\omega_i = \omega_p - \omega_s)$,  $\chi_{yyx}^{(p,2)} = \chi_{yyx}^{(2)}(\omega_p = \omega_i + \omega_s)$.  Index $\alpha$ in $\chi_{\alpha\beta\gamma}^{(2)}(\omega=\omega' \mp \omega'')$ corresponds to the polarization of the field at the mixing frequency $\omega$, and the index $\beta$ corresponds to the polarization of the field at a larger of the two frequencies $\omega',\omega''$. 

Using the nonlinear polarizations and boundary conditions for the fields, Eq.~(\ref{15}) becomes 
\begin{equation}
\label{40}
\frac{\partial \hat{a}_{s}}{\partial t}+v_{s}\frac{\partial \hat{a}_{s}}{\partial x}+(\gamma_{s}-\hat{G})\cdot\hat{a}_{s}=\hat{J} + \hat{F}_{s},
\end{equation}
where
\begin{align}
& \hat{J}=\Gamma\chi_{xyy}^{(s,2)}\hat{c}_{i}^{\dagger}\hat{c}_{p}, \;  
\Gamma=i\frac{2\pi \sqrt{\omega_{i}\omega_{p}}}{n_1^2} T_{i}T_{p}E_{s0}^{*},  \nonumber  \\
& \hat{G}= |\Gamma|^{2}\frac{n_{1}}{c}     \left(\frac{\chi_{xyy}^{(s,2)}\chi_{yyx}^{(i,2)*} \hat{c}_{p}^{\dagger}\hat{c}_{p}}{T_{i}\cos\theta_{1i}}  - \frac{\chi_{xyy}^{(s,2)}\chi_{yyx}^{(p,2)} \hat{c}_{i}^{\dagger}\hat{c}_{i}}{T_{p}\cos\theta_{1p}} 
\right).  \nonumber 
\end{align}
Here $T_{p,i} = 2 n_1 \cos\theta_{1p,i}/(n_1 \cos\theta_{1p,i} + n_2\cos\theta_{2p,i})$ are Fresnel transmission coefficients for s-polarized pump and idler fields with incidence angles $\theta_{1p,i}$ and refraction angles $\theta_{2p,i}$. Eq.~(\ref{40}) was derived neglecting the terms of the order $\alpha|\chi^{(2)}|^{2}$ and $|\chi^{(2)}|^{3}$ where $\alpha = e^2/\hbar c$.

The terms $\hat{J}$ and $\hat{G}$ in Eq.~(\ref{40}) include all possible three-wave mixing processes.  The term $\hat{J}$ describes difference frequency generation of surface plasmons in graphene by a bichromatic quantum field.  For classical fields this process has been predicted in \cite{5} and observed in \cite{hendry2015}.  The operator $\hat{G}$ describes the creation of plasmons by a parametric decay of the pump photons. 

The operator-valued Eq.~(\ref{40}) has a stationary solution given by 
\begin{align}
\label{43}
\hat{a}_{s}=\,&\exp\left[(\frac{\hat{G}-\gamma_{s}}{v_{s}})x\right]  \\ 
&\times\left(\hat{a}_{s}(0)+\int_{0}^{x}\exp\left[(\frac{\hat{G}-\gamma_{s}}{v_{s}})x'\right]^{-1}(\hat{J} + \hat{F}_s)  \frac{dx'}{v_{s}}\right) \nonumber ,
\end{align}
Here we will only deal with a coherent classical pump field at frequency $\omega_{p}$.  The field at the idler frequency $\omega_{i}$ is present only as a quantum and/or thermal noise.  In this case, and for $\hbar \omega_s \ll k_B T$, the term $\hat{J}$ can be neglected as compared to the Langevin noise term, and the operator $\hat{G}$ can be replaced by a c-number:
\begin{equation}
\label{44}
G  \approx |\Gamma|^{2}\frac{n_{1}}{c} \frac{\chi_{xyy}^{(s,2)}\chi_{yyx}^{(i,2)*}\left< \hat{c}_{p}^{\dagger}\hat{c}_{p} \right>}{T_{i}\cos\theta_{1i}}. 
\end{equation}

Taking the thermal noise as a boundary condition and taking into account Eqs.~(\ref{46}) and (\ref{43}) one can get 
\begin{align}
& \hat{a}_{s}^{\dagger}\hat{a}_{s}= \exp\left[2\frac{{\rm Re}[G]-\gamma_{s}}{v_{s}}x\right](\hat{a}_{s}^{\dagger}\hat{a}_{s})_{T}\label{47}\\
&\times\left[1+\frac{\gamma_{s}}{{\rm Re}[G]-\gamma_{s}}\left(1-\exp\left[-2\frac{{\rm Re}[G]-\gamma_{s}}{v_{s}}x\right]\right)\right], \nonumber
\end{align}
where the operator $(\hat{a}_{s}^{\dagger}\hat{a}_{s})_{T}$ corresponds to the thermal field and has an average value given by Eq.~(\ref{22}).
 Note that there is a  $1/v_s$ dependence in the gain factor  in Eq.~(\ref{47}) which describes the enhancement in the gain for slowly moving plasmons as compared to photons. 

From Eq.~(\ref{47}) one can obtain an important result, namely  the criterion for parametric instability:
\begin{align}
&{\rm Re}(\chi_{xyy}^{(s,2)}\chi_{yyx}^{(i,2)*})>0 \label{49}, \\
&{\rm Re}[G] \approx |\Gamma|^{2} \frac{{\rm Re}\left[\chi_{xyy}^{(s,2)}\chi_{yyx}^{(i,2)*} \right] I_p}{c^2 \hbar \omega_p} \frac{n_1^2}{T_{i}\cos\theta_{1i}} >\gamma_{s}, \label{50} 
\end{align}
where $I_{p}$ is the incident pump intensity.  

To calculate the magnitude of the parametric gain we need to substitute the components of the second-order susceptibility tensor. Their derivation is straightforward but cumbersome, so we keep it in the Supplemental Material; see Eq.~(S10). Their salient feature  is the presence of resonances when one of the three frequencies involved in three-wave mixing is close to $2 \epsilon_F = 2 \hbar v_F k_F$. This is a weaker resonance than the one that would exist in coupled quantum wells \cite{belkin2007} where $\chi^{(2)}$ would scale as a product of two Lorentzians. Still, it enhances the value of $\chi^{(2)}$ by a factor of $\omega/\gamma$. A similar resonance exists in the third-order nonlinear response of graphene \cite{mikhailovPRB2014}. Close to resonance  one has to include the imaginary part of the frequency which describes the decay rate of the optical or plasmon polarization. We will take the same values for the imaginary part $\gamma$ for both pump and idler frequencies. Furthermore we assume $\omega_{p,i} \gg \omega_s \gg \gamma_s$ and consider strongly degenerate graphene with a frequency of the pump field close to $2 v_F k_F$. In addition to resonant enhancement of the nonlinearity, this eliminates interband absorption losses for the plasmons and reduces electron scattering. Under these conditions, when $|\omega_p - 2 v_F k_F| < \gamma$, we obtain
\begin{equation}
\label{pumpres}
\chi^{(s,2)}_{xyy} = \chi_{yyx}^{(i,2)*} \approx   \frac{3 e^3 v_F^2}{16 \pi \hbar^2} \frac{ q_p}{ \omega_i \omega_s^2 \gamma}. 
\end{equation}

\begin{figure}[htb]
\centerline{
\includegraphics[width=7cm]{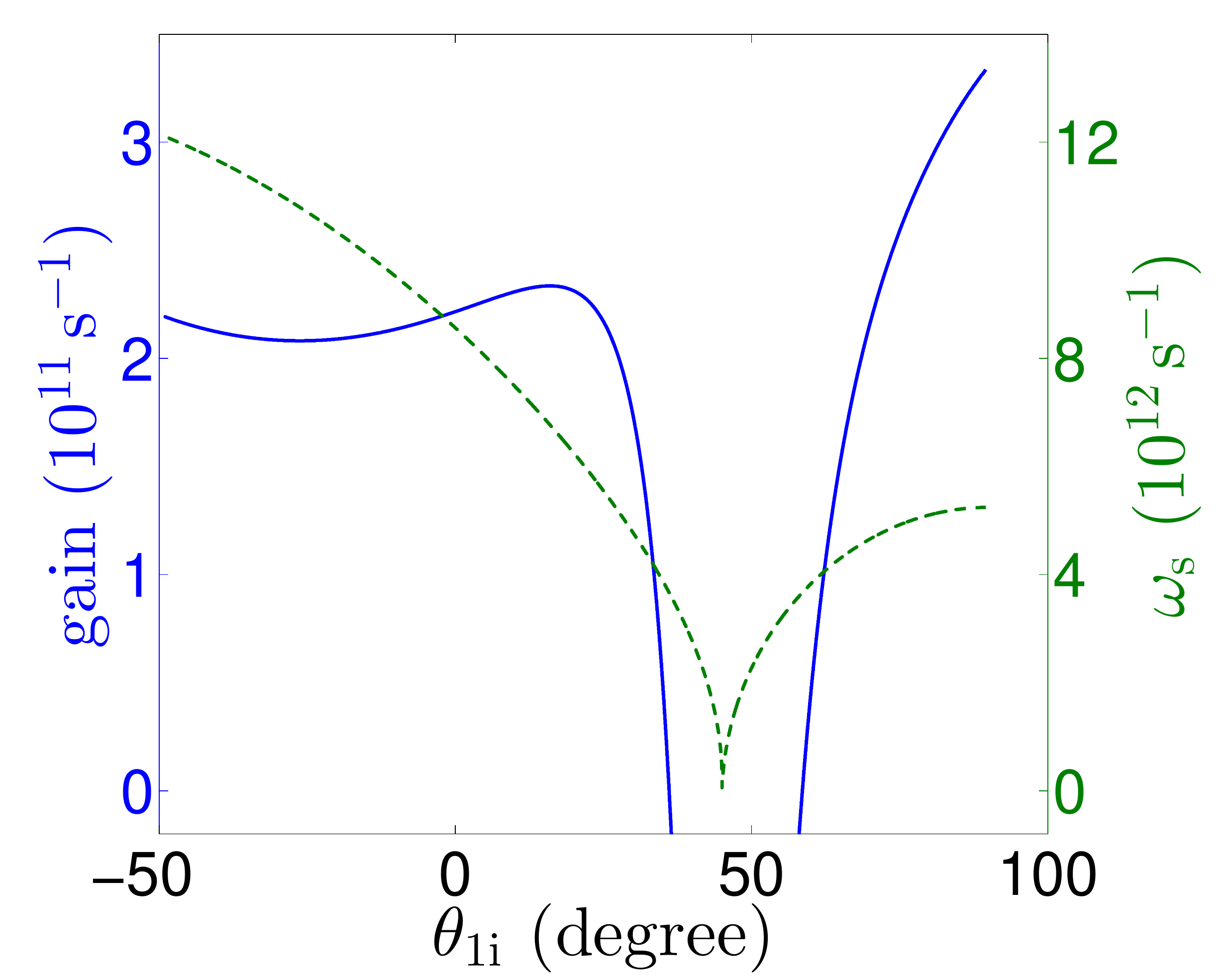}}
  \caption{The gain Re$[G]$ (solid blue line) and  the plasmon frequency corresponding to phase matching conditions (green dashed line) as a function of the angle $\theta_{1i}$ between the direction of the idler wave vector in medium 1 and the normal. }
  \end{figure}

\begin{figure}[htb]
\centerline{
\includegraphics[width=6cm]{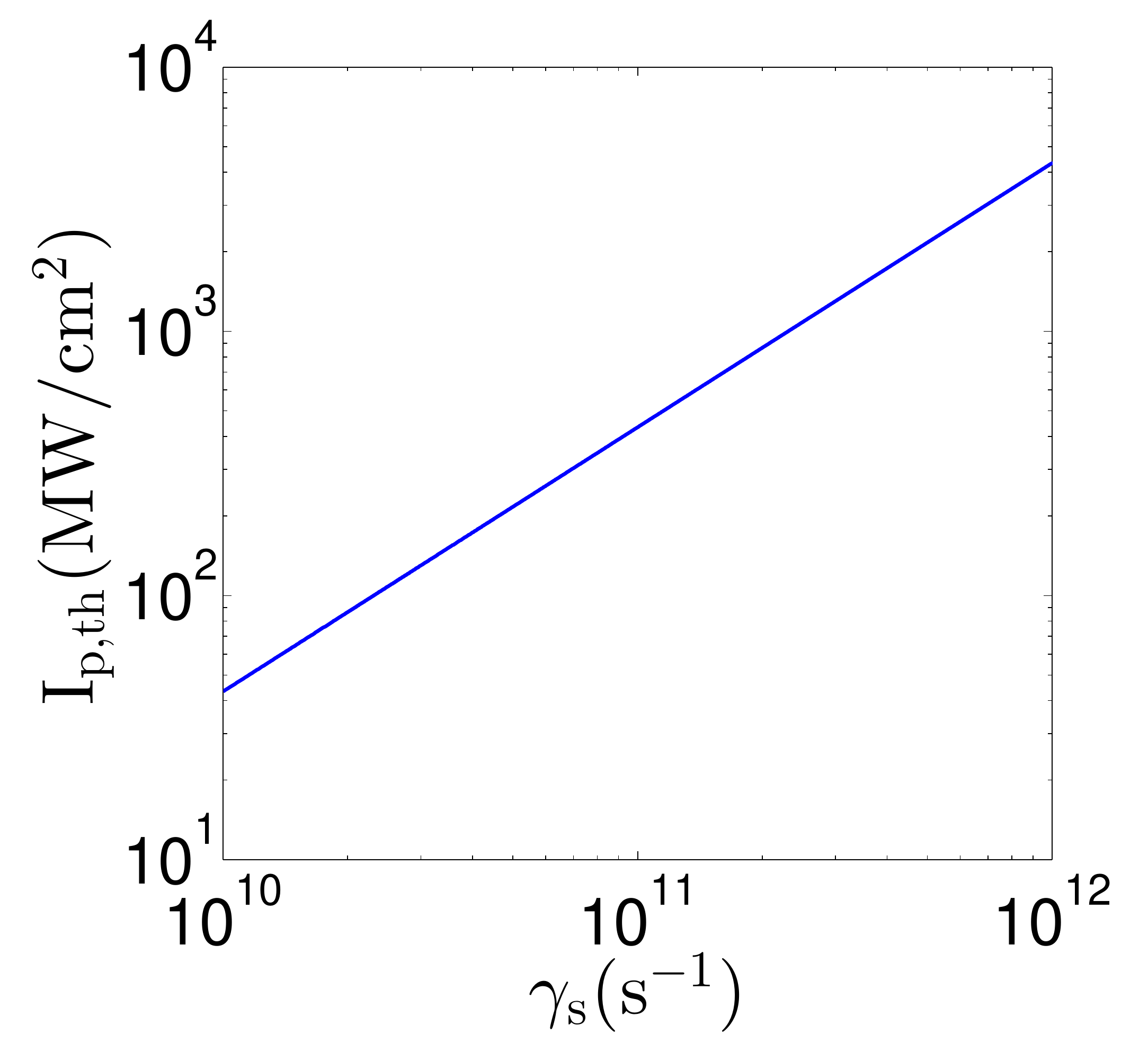}}
  \caption{The pumping intensity $I_p$ needed to reach the parametric instability threshold, Re$[G] = \gamma_s$ in Eq.~(\ref{50}), as a function of the plasmon decay rate $\gamma_s$.}
  \end{figure}

Figure 2 shows the gain (left-hand side of Eq.~(\ref{50})) and the plasmon frequency corresponding to phase matching conditions as a function of the emission angle of the idler field $\theta_{1i}$. Negative angles correspond to negative projections of $q_i$. For the plot we took $n_1 = 1$ and $n_2 = 2$, $\gamma_{p,i} = 10^{12}$ s$^{-1}$, the pump beam at a $10$-$\mu$m wavelength and incidence angle of $\pi/4$. The pumping intensity was assumed to be $I_p = 1$ GW/cm$^2$, which corresponds to intensities in the experiment \cite{hendry2015}. The gain is only weakly dependent on the idler emission angle except for a narrow range around $\theta_{1i} = \pi/4$ where $q_p \simeq q_i$ and therefore $q_s, \omega_s \rightarrow 0$. In this range the gain becomes negative; however, the approximation $\omega_s \gg \gamma_s$ becomes invalid, so this case requires a separate investigation.

In Fig.~3 we show the pumping intensity $I_p$ needed to reach the parametric instability threshold, Re$[G] = \gamma_s$, as a function of plasmon decay rate $\gamma_s$, for the same numerical parameters as in Fig.~2 and  for the idler emitted at $\theta_{1i} = 20$ degrees.  In this case the phase matching condition is satisfied when the plasmon frequency $\omega_s/2\pi$ is equal to 1 THz (see Fig.~2).  

The magnitude of the gain can be further increased by non-Bernal stacking of multiple graphene layers, which will reduce the threshold intensity. 

Low-energy surface states of a 3D topological insulator Bi$_2$Se$_3$ are massless 2D Dirac fermions described by the effective Hamiltonian $H = v_{F} (\vec{\sigma} \times \vec{p})_z$ \cite{parameters}, where $v_F$ is two times smaller than in graphene. The states have different chirality as compared to those in graphene but the same matrix elements of the interaction Hamiltonian and the same structure of the optical response. Proceeding in the same way, one can show that the parametric gain for a thin Bi$_2$Se$_3$ film (i.e.~two surfaces) will have a magnitude lower by a factor of $\sim 2^6$ due to a two times lower $v_F$ and two times lower degeneracy. \\

\section{Generated idler field flux} 

The outgoing flux of idler photons generated as a result of parametric decay of the pump carries information on the intensity of generated surface plasmon field and the surface nonlinearity.  To calculate the average flux of the idler photons on the detector we calculate first the operator of the idler field generated by the nonlinear current. Using Eqs.~(\ref{polar}) and standard boundary conditions, we arrive at
\begin{equation}
\delta \bm{\hat{E}}_i = \bm{y}_{0} E_{i0} \hat{c}_i^{(2)} {\rm e}^{i k_i z + i q_i x -i\omega_{i}t}+{\rm H.c.}, \, \hat{c}_i^{(2)} = \frac{n_1 \Gamma \chi_{yyx}^{(i,2)}}{c \cos\theta_{1i}} \hat{a}_s^{\dagger} \hat{c}_p. \nonumber 
\end{equation}
One also needs to calculate the average value of the generated number of plasmon quanta $\left<\hat{a}_{s}^{\dagger}\hat{a}_{s}\right>_{D}$, which is collected by the detector from length $L_{x}$. Using Eq.~(\ref{47}) we obtain
\begin{align}
\left<\hat{a}_{s}^{\dagger}\hat{a}_{s}\right>_{D} & =\frac{1}{L_{x}}\int_{0}^{x}\left<\hat{a}_{s}^{\dagger}\hat{a}_{s}\right>dx\approx\left<\hat{a}_{s}^{\dagger}\hat{a}_{s}\right>_{T}\frac{e^{\Xi}-1}{\Xi};  \label{48}\\
\Xi & =2\frac{{\rm Re}[G]-\gamma_{s}}{v_{s}}L_{x}.
\nonumber
\end{align}
The resulting average flux of the idler photons  on the detector of transverse area $A_{D}$ is given by 
\begin{equation}
\nonumber 
\left<\hat{\Pi}_{i}^{(2)}\right> = \frac{n_1^2 |\Gamma|^{2}|\chi_{yyx}^{(i,2)}|^2 I_p L_x  \Delta \omega }{2 \pi c^{2} v_s \hbar \omega_p \cos\theta_{1i} } \left( \frac{e^{\Xi}-1}{\Xi} N_T(\omega_s) +1 \right),
\end{equation}
where $\hat{\Pi}_{i}^{(2)} = \displaystyle \frac{c A_D}{n_1} \hat{c}_{i}^{\dagger}\hat{c}_{i}$. 

Close to the parametric instability threshold, when  $\displaystyle \frac{e^{\Xi}-1}{\Xi} \sim 1$, the idler photon flux is
\begin{equation}
\label{56}
\left<\hat{\Pi}_{i}^{(2)}\right> \sim \frac{\Delta\omega\gamma_{s}L_{x}}{2\pi v_{s}}N_{T}(\omega_{s}), 
\end{equation}
i.e.~ it is of the order of the thermal flux at a much lower surface plasmon frequency collected from the length equal to the plasmon decay length, $L_x \sim v_s/\gamma_s$. \\

\section{Plasmon-photon entanglement}

The total idler field propagating away from the graphene layer to the detector consists of the reflected noise field and the generated parametric field calculated above: 
\begin{equation}
\label{58}
\hat{c}_{r}\approx R_{i}\hat{c}_{i} + \frac{n_1 \Gamma \chi_{yyx}^{(i,2)}}{c \cos\theta_{1i}} \hat{a}_s^{\dagger} \hat{c}_p,
\end{equation}
where
$ R_{i}= \displaystyle \frac{n_{1}\cos\theta_{1i}-n_{2}\cos\theta_{2i}}{n_{1}\cos\theta_{1i}+n_{2}\cos\theta_{2i}}$ is the Fresnel reflection coefficient for the S-polarized field. Here we neglected thermal noise at high frequencies,  $N_T(\omega_{p,i})$, and absorption in monolayer graphene $\sim \pi \alpha$. Calculating quantum-mechanical averages of the quantities quadratic with respect to the reflected field, it is easy to see that Eq.~(\ref{58}) corresponds to an entangled plasmon-photon state. In particular, for a classical pump field  and an initial state in the form $\Psi_0 = \left| 0_i \right> \left|0_s \right>$, one can show \cite{14} that parametric decay leads to the state of the kind $\Psi=\alpha \left|0_{i}\right>\left|0_{s}\right>+ \beta \left|1_{i}\right>\left|1_{s}\right>$. 

In general, the calculations of quantum-mechanical averages of any physical quantities are much easier to perform in the Heisenberg picture using Eq.~(\ref{58}) for Heisenberg operators, without converting to the Schr\"{o}dinger picture.  In particular, it is obvious from Eq.~(\ref{58}) that for a given spectrum of the pump field any physical observable for a surface plasmon field can be related to a corresponding observable for the idler field at frequency $\omega_{i}$.  For example, if the pump field spectrum is much narrower than the spectrum of the plasmon fluctuations then the spectrum of surface plasmons is related to the spectrum of idler photons.

In conclusion, we showed the feasibility of stimulated parametric decay of photons of a strong laser pump obliquely incident on graphene. We calculated the flux of surface plasmons and idler photons generated by parametric decay of the pump, and demonstrated their entanglement.  \\

 \section{Supplemental Material}

\subsection{Second-order nonlinear susceptibility in graphene}

In this Supplemental Material we provide the general expressions for the components of the second-order susceptibility tensor that are relevant for the parametric three-wave mixing in graphene. 

The Hamiltonian of graphene near the Dirac point $\bm{K}$ is
\begin{eqnarray}
H = v_{F} \bm{\sigma} \cdot \hat{\bm{p}}
 = v_{F} \begin{pmatrix}
 0 & \hat{p}_x-i \hat{p}_y \\
 \hat{p}_x+i \hat{p}_y & 0
  \end{pmatrix},
\end{eqnarray}  
where $\hat{\bm{p}}$ is the momentum operator relative to $\bm{K}$ and  $\bm{\sigma}$ is a 2D vector of Pauli matrices. The eigenenergies are $\epsilon_{\pm}(\bm{k}) = \pm\hbar v_F k$, and eigenstates are
\begin{eqnarray}
\langle \bm{r}_{\|} | s, \bm{k} \rangle =\frac{1}{\sqrt{2 A}}\exp(i\bm{k}\cdot \bm{r}_{\|} )\left(\begin{array}{c}
s\\
e^{i\phi(\bm{k})}
\end{array}\right) ,
\label{Eq:wave_func_nomagnetic}
\end{eqnarray}
where $s = 1$ for conduction band, $s = -1$ for valence band, $A$ is the area of graphene, and $\phi(\bm{k})$ is the angle of the wave vector $\bm{k}$ with the $x$-axis.

 The interaction Hamiltonian between graphene and the optical field which has an in-plane component of the electric field can be written as
 \begin{eqnarray}
 \hat{H}^{op}_{int} = v_F \frac{e}{c} \bm{\sigma} \cdot \bm{A} = \frac{e}{c} \hat{\bm{v}} \cdot \bm{A} ,
 \end{eqnarray}
where $\hat{\bm{v}}$ = $v_F \bm{\sigma}$ is the velocity operator, and $\bm{A}$ is the vector potential of the optical field, which is related to the electric field by $\bm{E}$ = $(-1/c) \partial\bm{A} / \partial t$. Using this Hamiltonian, the evolution equation for the density matrix is given by
\begin{align}
i\hbar \frac{\partial}{\partial t} \rho_{mn} &= (\epsilon_m - \epsilon_n) \rho_{mn} + \frac{e}{c} (\hat{\bm{v}} \cdot \bm{A})_{mn} (\rho_{nn} - \rho_{mm} ) \nonumber \\
&+ \frac{e}{c} \sum_{l\neq m,n} \left( (\hat{\bm{v}} \cdot \bm{A})_{ml} \rho_{ln} - \rho_{ml} (\bm{v}\cdot \bm{A})_{ln} \right) ,
\end{align} 
where both linear and nonlinear effects are included. We calculate the field-induced current in second order with respect to the optical field, as a quantum-mechanical average of the current operator  $\hat{\bm{j}} = -e v_F \bm{\sigma}$ with the density matrix.

We will seek the response at the sum frequency $\omega_1+\omega_2$ to the bichromatic optical field with in-plane electric fields at frequencies $\omega_{1,2}$ directed along unit vectors $\bm{\eta}_{1,2}$ : 
\begin{align}
\bm{A} = \frac{1}{2} \bm{\eta}_1 A(\omega_1) e^{i(\bm{q}_1\cdot\bm{r}_{\|} - \omega_1 t)} + \frac{1}{2} \bm{\eta}_2 A(\omega_2) e^{i(\bm{q}_2\cdot\bm{r}_{\|} - \omega_2 t)} + \mathrm{c.c.}
\end{align}
The result will be applicable to the difference frequency process by choosing either positive or negative frequencies, with the corresponding change in  $\bm{q}$ for a given $\omega$. The second-order density matrix elements at the sum frequency $\omega_1+\omega_2$ are evaluated to be
\begin{align}
&\phantom{{}={}}\rho_{mn}^{(2)}(\omega_1+\omega_2)  = 
\frac{1}{2} \left(\frac{e}{c}\right)^2 
\frac{ A(\omega_1) A(\omega_2) }{\hbar(\omega_1+\omega_2)-(\epsilon_m-\epsilon_n)}  \nonumber \\
&\times \sum_{l\neq m,n} \left( (\hat{\bm{v}} \cdot \bm{\eta}_1) e^{i\bm{q}_1 \cdot \bm{r}_{\|}} \right)_{ml} 
\left( (\hat{\bm{v}} \cdot \bm{\eta}_2) e^{i\bm{q}_2 \cdot \bm{r}_{\|}} \right)_{ln}  \nonumber \\
&\times \left[
\frac{ (\rho_{nn}-\rho_{ll}) }{ \hbar\omega_2 - (\epsilon_l - \epsilon_n) }
- \frac{ (\rho_{ll}-\rho_{mm}) }{  \hbar\omega_1 - (\epsilon_m - \epsilon_l) }
\right] \nonumber \\
&+ \left\{ 1 \leftrightarrow 2 \right\} .
\end{align}

The matrix elements entering the above expression are given by 
\begin{align}
&\left( (\hat{\bm{v}} \cdot \bm{\eta}) e^{i\bm{q} \cdot \bm{r}_{\|}} \right)_{mn} = \frac{1}{2} v_F \left[ (\eta_x - i\eta_y) s_m e^{i\phi_n} \right.  \nonumber \\ 
& \left. + (\eta_x + i\eta_y) s_n e^{-i\phi_m} \right] \delta_{\bm{k}_m,\bm{k}_n+\bm{q}}.
\end{align}

The average of the corresponding Fourier harmonic of the induced current with the density matrix can be calculated as
\begin{align}
\bm{J}^{(2)}(\omega_1+\omega_2) = - e \sum_{mn} \left(\hat{\bm{v}} e^{-i(\bm{q}_1+\bm{q}_2)\cdot\bm{r}_{\|}}\right)_{nm} \rho_{mn}^{(2)}(\omega_1+\omega_2) .
\end{align}

Next, we transform from summation to integration over $\bm{k}$-states, introduce the corresponding occupation numbers $f(s,\bm{k})$ of the momentum states in each band, apply the momentum conservation in a three-wave mixing process, and take into account spin and valley degeneracy. The result is 
\begin{align}
&\phantom{{}={}} \bm{J}^{(2)}(\omega_1+\omega_2) 
= - \frac{e^3 v_F^2}{16 \pi^2 c^2 \hbar^2}  A(\omega_1) A(\omega_2)   \nonumber \\ 
& \sum_{s_m,s_n,s_l} \int d^2\bm{k} 
\frac{1}
{
(\omega_1+\omega_2)- v_F (s_m | \bm{k}+\bm{q}_1 |  - s_n |\bm{k}-\bm{q}_2| )
} \nonumber \\
&\times \left[
\frac{ f(s_n,|\bm{k}-\bm{q}_2|) - f(s_l,|\bm{k}|) }
{\omega_2 -  v_F (s_l |\bm{k}| - s_n |\bm{k}-\bm{q}_2|)} \right.  \nonumber \\ 
& \left. -
\frac{ f(s_l,|\bm{k}|) - f(s_m,|\bm{k}+\bm{q}_1|) }
{ \omega_1 -  v_F (s_m |\bm{k}+\bm{q}_1| - s_l |\bm{k}|) }
\right] \nonumber \\
&\times \left[ (\eta_{1x} - i\eta_{1y}) s_m e^{i\phi(\bm{k})} + (\eta_{1x} + i\eta_{1y}) s_l e^{-i\phi(\bm{k}+\bm{q}_1)} \right]  \nonumber \\
&\times \left[ (\eta_{2x} - i\eta_{2y}) s_l e^{i\phi(\bm{k}-\bm{q}_2)} + (\eta_{2x} + i\eta_{2y}) s_n e^{-i\phi(\bm{k})} \right] \nonumber \\
&\times \left[ (\hat{x} + i \hat{y}) s_m e^{-i\phi(\bm{k}-\bm{q}_2)} + (\hat{x} - i \hat{y}) s_n e^{i\phi(\bm{k}+\bm{q}_1)} \right]  \nonumber \\
&+ \left\{ 1 \leftrightarrow 2 \right\} .
\label{Eq:J_2ndorder_general} 
\end{align}

This equation can be integrated numerically for any given geometry of incident fields and electron distribution. We consider the limit of the Fermi distribution with a strong degeneracy, direct all in-plane photon wave vectors along x-axis,  and expand the integrand in Eq.~(\ref{Eq:J_2ndorder_general}) in powers of $q_1, q_2$. The integral over the term of zeroth-order in $q$ vanishes, as expected from symmetry. We will keep the terms linear in $q$.  Also we have to evaluate separately the intraband contribution $s_l=s_m=s_n$ and all types of mixed interband-intraband contributions:  $s_m = s_n = -s_l$, $s_m = s_l = -s_n$, and $s_n = s_l = -s_m$. 
Here we give only the component of the second-order nonlinear conductivity tensor which gives the main contribution to the signal:
\begin{align}
&\phantom{{}={}}\sigma^{(2)}_{xyy}(\omega_1+\omega_2;\omega_1,\omega_2)  = 
- s(\epsilon_F) \frac{e^3 v_F^2}{2 \pi \hbar^2}  \frac{1}{\omega_1^2 \omega_2^2 (\omega_1+\omega_2)} \nonumber \\ &\times \frac{1}
{(\omega_1^2 - 4 v_F^2 k_F^2) (\omega_2^2 - 4 v_F^2 k_F^2) ((\omega_1 + \omega_2)^2 - 4 v_F^2 k_F^2 )} \nonumber \\
&\times \left[ 4(v_F k_F)^2 \omega_1 \omega_2 (\omega_1+\omega_2)^2 (q_1 \omega_2^2 + q_2 \omega_1^2) \right. \nonumber \\
&\phantom{{}={}} + 4 (v_F k_F)^4 (q_1 \omega_2^4 - (6 q_1 + 4 q_2) \omega_1 \omega_2^3 \nonumber \\
&- 8 (q_1+q_2) \omega_1^2 \omega_2^2 - (4 q_1 + 6 q_2) \omega_1^3 \omega_2 + q_2 \omega_1^4 )  \nonumber \\
&\phantom{{}={}} - \left. 16 (v_F k_F)^6 ( q_1 \omega_2 ( \omega_2 - 2\omega_1) + q_2 \omega_1 ( \omega_1 - 2  \omega_2)) \right].
\label{sigmaxyy}
\end{align}
Here $s(\epsilon_F) = \pm 1$ depending on whether the Fermi level is in the conduction or valence band. 
The result for the difference frequency can be obtained from Eq.~(\ref{sigmaxyy}) by flipping the sign of $\omega_2$ and $q_2$. 

After converting the nonlinear conductivity to the nonlinear susceptibility according to  
$$ \chi^{(2)}_{ijk}(\omega_1+\omega_2;\omega_1,\omega_2) = \frac{i \sigma_{ijk}^{(2)}(\omega_1+\omega_2;\omega_1,\omega_2)}{\omega_1 + \omega_2 },
$$
one can verify that in the absence of dissipation all components of the nonlinear susceptibility tensor that we calculated satisfy permutation relations originated from symmetry properties; see e.g. Ch.~2.9 in \cite{keldysh}:
\begin{align}
&\chi^{(2)}_{ijk}(\omega_3=\omega_1+\omega_2) = \chi^{(2)}_{jik}(-\omega_1=-\omega_3+\omega_2) \nonumber \\
& = \chi^{(2)}_{kji}(-\omega_2=-\omega_3+\omega_1),  \label{permut}
\end{align}
 where in-plane wave vectors have to be permuted together with frequencies. 
  
 The second-order response goes to zero when the Fermi energy $\epsilon_F$ goes to zero, and is maximized when one of 
the three frequencies involved in three-wave mixing is close to $2 \epsilon_F/\hbar = 2  v_F k_F$. Close to resonance with $2 \epsilon_F/\hbar$ one has to include the imaginary part of the frequency which comes from the omitted relaxation term $-\gamma \rho_{mn}$ in the density-matrix equations. This amounts to substituting 
$\omega_1 \rightarrow \omega_1 + i \gamma_1$, $\omega_2 \rightarrow \omega_2 + i \gamma_2$, $\omega_1 + \omega_2 \rightarrow \omega_1 + \omega_2 + i \gamma_3$. Note that if we flip the sign of $\omega_2$ the sign of $+ i \gamma_2$ remains the same. If dissipation is included, one cannot use permutation relations Eq.~(\ref{permut}) and has to evaluate each component of  $\chi^{(2)}_{ijk}$ independently. 

\subsection{A coupled oscillators model for the parametric gain}

The instability condition Eq.~(\ref{50})  can be easily interpreted and understood within the  classical model of two parametrically coupled oscillators.  Consider a classical pump beam of amplitude $E_{p}$ and $\omega_{p}$ incident on a nonlinear 2D layer in vacuum.  The pump field decays into a surface plasmon field within a unit area $A_{s}=1$ and an idler photon field at frequency $\omega_{i}$ within a volume of a cylinder of length $l$ oriented at an angle $\theta_{i}$ with respect to the normal to area $A_{s}$.  In this mean-field zero-dimensional (0D) model one can derive the following coupled differential equations for the complex amplitudes of the plasmon and idler fields:
\begin{align}
&\frac{\partial E_{s}}{\partial t}+\gamma_{s}E_{s}=i\zeta_{s}E_{p}E_{i}^{*}\label{osc1},\\
&\frac{\partial E_{i}^{*}}{\partial t}+\gamma_{i}E_{i}^{*}=-i\zeta_{i}^{*}E_{p}^{*}E_{s}\label{osc2},
\end{align}  
where
\begin{align}
&\zeta_{s}=\frac{1}{2}\chi_{xyy}^{(s,2)} \left[{\rm Re}\left(\frac{\partial \chi_{s}}{\partial \omega}\right)\right]^{-1}, \nonumber \\
& \zeta_{i}=\frac{\pi}{l\cos\theta_{i}}\omega_{i}\chi_{yyx}^{(i,2)*},\nonumber
\end{align}
$\gamma_{i}=c/l$ is the effective decay rate of the idler field in the 0D model.  Equations (\ref{osc1}) and (\ref{osc2}) have an exponentially growing solution for both parametrically coupled waves \cite{shen} if ${\rm Re}(\zeta_{s}\zeta_{i}^{*})|E_p|^2 > \gamma_{s}\gamma_{i}$, which coincides with Eq.~(\ref{50}), if we use Eq.~(\ref{9}) and assume $n_{2}=n_{1}= 1$.

\end{document}